\documentclass[showpacs,english,aps,prc,twocolumn,floatfix]{revtex4}
\usepackage[T1]{fontenc}
\usepackage[latin1]{inputenc}
\usepackage{graphicx}
\usepackage{amssymb}
\usepackage{verbatim}
\usepackage{epic,eepic}
\usepackage{babel}

\newcommand{\ba}{\begin{eqnarray}}
\newcommand{\ea}{\end{eqnarray}}

\begin{document}
\pagestyle{plain}

\title{New developments in nuclear supersymmetry
\footnote{Invited talk at XXXII Symposium on Nuclear Physics, 
Hacienda Cocoyoc, Morelos, Mexico, January 5-8, 2009}}

\author{R. Bijker}
\email{bijker@nucleares.unam.mx}
\affiliation{Instituto de Ciencias Nucleares, 
Universidad Nacional Aut\'onoma de M\'exico, 
A.P. 70-543, 04510 M\'exico, D.F., M\'exico}

\author{J. Barea}
\email{jose.barea@iem.cfmac.csic.es}
\affiliation{Facultad de F{\'{\i}}sica, Universidad de Sevilla, 
Avda. Reina Mercedes s/n, E-4012 Sevilla, Espa\~{n}a}

\author{A. Frank}
\email{frank@nucleares.unam.mx}
\affiliation{Instituto de Ciencias Nucleares, 
Universidad Nacional Aut\'onoma de M\'exico, 
A.P. 70-543, 04510 M\'exico, D.F., M\'exico}

\date{January 14, 2009}

\begin{abstract}
We discuss several new developments in nuclear supersymmetry, in particular the identification 
of a new supersymmetric quartet of nuclei in the $A \sim 190$ mass region consisting of 
the $^{192,193}$Os and $^{193,194}$Ir nuclei, and a study of correlations between 
different transfer reactions.

\

\end{abstract}


\maketitle

\section{Introduction}

Nuclear structure physics has seen an impressive progress in the development 
of {\em ab initio} methods (no-core shell model, Green's Function Monte Carlo, 
Coupled Clusters, ...), mean-field techniques and effective field theories for 
which the ultimate goal is {\em an exact treatment of nuclei utilizing the 
fundamental interactions between nucleons} \cite{bluebook}. All involve large 
scale calculations and therefore rely heavily on the available computing power 
and the development of efficient algorithms to obtain the desired results. 

A different, complementary, approach is that of symmetries and algebraic methods. 
Rather than trying to solve the complex nuclear many-body problem numerically, the 
aim is to identify effective degrees of freedom, develop schematic models based upon 
these degrees of freedom and study their solutions by means of symmetries, etc. Aside 
from their esthetic appeal, symmetries provide energy formula, selection rules 
and closed expressions for electromagnetic transition rates and transfer strengths 
which can be used as benchmarks to study and interpret the experimental data, 
even if these symmetries may be valid only approximately. Historically, 
symmetries have played an important role in nuclear physics. 
Examples are isospin symmetry, the Wigner supermultiplet theory, special solutions 
to the Bohr Hamiltonian, the Elliott model, pseudo-spin symmetries and the dynamical 
symmetries and supersymmetries of the IBM and its extensions. 

The purpose of this contribution is to discuss several new developments in nuclear 
supersymmetry, in particular evidence for the existence of a new supersymmetric 
quartet in the $A \sim 190$ mass region, consisting of the $^{192,193}$Os and $^{193,194}$Ir 
nuclei, and correlations between different one- and two-nucleon transfer reactions. 

\section{Nuclear supersymmetry}

Nuclear supersymmetry is a composite-particle phenomenon that should not be confused 
with fundamental supersymmetry, as used in particle physics and quantum field theory  
where it is postulated as a generalization of the Lorentz-Poincar\'e invariance as a 
fundamental symmetry of Nature and predicts the existence of supersymmetric particles, 
such as the photino and the selectron, for which experimental evidence is yet 
to be found. If experiments about to start at the LHC at CERN find evidence of supersymmetric 
particles, supersymmetry would be badly broken, as their masses must be much higher than 
those of their normal partners. In contrast to particle physics, nuclear supersymmetry has 
been verified experimentally.

Dynamical supersymmetries were introduced in nuclear physics in 
the context of the Interacting Boson Model (IBM) and its extensions \cite{FI}.  
The IBM describes collective excitations in even-even nuclei in 
terms of a system of interacting monopole ($s^{\dagger}$) and quadrupole 
($d^{\dagger}$) bosons \cite{IBM}. The bosons are associated with the number of 
correlated proton and neutron pairs, and hence the number of bosons $N$ is 
half the number of valence nucleons.  
For odd-mass nuclei the IBM was extended to include single-particle
degrees of freedom \cite{olaf}. The ensuing Interacting Boson-Fermion Model 
(IBFM) has as its building blocks $N$ bosons with $l=0,2$ and $M=1$ fermion 
($a_j^{\dagger}$) with $j=j_1,j_2,\dots$ \cite{IBFM}. The IBM and IBFM can 
be unified into a supersymmetry (SUSY) $U(6/\Omega) \supset U(6) \otimes U(\Omega)$
where $\Omega=\sum_j (2j+1)$ is the dimension of the fermion space \cite{FI} . 
In this framework, even-even and odd-even nuclei form the members of a 
supermultiplet which is characterized by ${\cal N}=N+M$, 
{\em i.e.} the total number of bosons and fermions.  
Supersymmetry distinguishes itself from other symmetries in that it includes, 
in addition to transformations among fermions and among bosons, also 
transformations that change a boson into a fermion and {\em vice versa}.

The concept of nuclear SUSY was extended in 1985 to include the neutron-proton 
degree of freedom \cite{quartet}. In this case, a supermultiplet consists of an 
even-even, an odd-proton, an odd-neutron and an odd-odd nucleus. The first 
experimental evidence of a supersymmetric quartet was found in the $A \sim 190$ 
mass region in the $^{194,195}$Pt and $^{195,196}$Au nuclei as an example of the 
$U(6/12)_{\nu} \otimes U(6/4)_{\pi}$ supersymmetry \cite{metz,groeger,wirth,barea1,barea2}, 
in which the odd neutron is allowed to occupy the $3p_{1/2}$, $3p_{3/2}$ and $2f_{5/2}$ 
orbits of the 82-126 shell, and the odd proton the $2d_{3/2}$ orbit of the 50-82 shell. 
This mass region is a particularly complex one, displaying transitional behavior such 
as prolate-oblate deformed shapes, $\gamma$-unstability, triaxial deformation and/or 
coexistence of different configurations which present a daunting challenge to nuclear 
structure models. Nevertheless, despite its complexity, the $A \sim 190$ mass region 
has been a rich source of empirical evidence for the existence of dynamical symmetries 
in nuclei both for even-even, odd-proton, odd-neutron and odd-odd nuclei, as well as 
supersymmetric pairs \cite{FI,baha} and quartets of nuclei \cite{quartet,metz}. 

Recently, the structure of the odd-odd nucleus $^{194}$Ir was investigated by a series 
of transfer and neutron capture reactions \cite{balodis}. 
The odd-odd nucleus $^{194}$Ir differs from $^{196}$Au by two protons, the 
number of neutrons being the same. The latter is crucial, since the dominant 
interaction between the odd neutron and the core nucleus is of quadrupole type, 
which arises from a more general interaction in the IBFM for very special values 
of the occupation probabilities of the $3p_{1/2}$, $3p_{3/2}$ and $2f_{5/2}$ 
orbits, {\em i.e.} to the location of the Fermi surface for the neutron orbits 
\cite{bijker}. This situation is satisfied to a good approximation by the 
$^{195}$Pt and $^{196}$Au nuclei which both have the 117 neutrons. The same is 
expected to hold for the isotones $^{193}$Os and $^{194}$Ir. For this reason, 
it is reasonable to expect the odd-odd nucleus $^{194}$Ir to provide another example 
of a dynamical symmetry in odd-odd nuclei. Strictly speaking, a 
dynamical supersymmetry involves the simultaneous study of pairs or quartets of 
nuclei that make up a supermultiplet. 

\subsection{Energies}

In general, a dynamical (super)symmetry arises whenever the Hamiltonian is expressed in 
terms of the Casimir invariants of the subgroups in a group chain. The relevant subgroup 
chain of the $U(6/12)_{\nu} \otimes U(6/4)_{\pi}$ supersymmetry is given by \cite{quartet}
\begin{eqnarray}
U(6/12)_{\nu} &\otimes& U(6/4)_{\pi} 
\nonumber\\
&\supset& U^{B_{\nu}}(6) \otimes U^{F_{\nu}}(12) \otimes
U^{B_{\pi}}(6) \otimes U^{F_{\pi}}(4)
\nonumber\\
&\supset& U^B(6) \otimes U^{F_{\nu}}(6) \otimes U^{F_{\nu}}(2) \otimes
U^{F_{\pi}}(4)
\nonumber\\
&\supset& U^{BF_{\nu}}(6) \otimes U^{F_{\nu}}(2) \otimes U^{F_{\pi}}(4)
\nonumber\\
&\supset& SO^{BF_{\nu}}(6) \otimes U^{F_{\nu}}(2) \otimes SU^{F_{\pi}}(4)
\nonumber\\
&\supset& Spin(6) \otimes U^{F_{\nu}}(2)
\nonumber\\
&\supset& Spin(5) \otimes U^{F_{\nu}}(2)
\nonumber\\
&\supset& Spin(3) \otimes SU^{F_{\nu}}(2)
\nonumber\\
&\supset& SU(2) ~.
\label{chain}
\end{eqnarray}
In this case, the Hamiltonian
\begin{eqnarray}
H &=& A \, C_{2U^{BF_{\nu}}(6)} + B \, C_{2SO^{BF_{\nu}}(6)}
+ B' \, C_{2Spin(6)}
\nonumber\\
&& + C \, C_{2Spin(5)} + E \, C_{2Spin(3)}
+ F \, C_{2SU(2)} ~,
\end{eqnarray}
describes the excitation spectra of the quartet of nuclei.
Here we have neglected terms that only contribute to binding energies.
The energy spectra of the four nuclei belonging to the supersymmetric 
quartet are described simultaneously by a single energy formula in terms 
of the eigenvalues of the Casimir operators
\begin{eqnarray}
E &=& A \left[ N_1(N_1+5)+N_2(N_2+3)+N_1(N_1+1) \right] 
\nonumber\\
&& + B  \left[ \Sigma_1(\Sigma_1+4)+\Sigma_2(\Sigma_2+2)+\Sigma_3^2 \right] 
\nonumber\\
&& + B' \left[ \sigma_1(\sigma_1+4)+\sigma_2(\sigma_2+2)+\sigma_3^2 \right] 
\nonumber\\
&& + C \left[ \tau_1(\tau_1+3)+\tau_2(\tau_2+1) \right] 
\nonumber\\
&& + D \, L(L+1) + E \, J(J+1) ~.
\label{npsusy}
\end{eqnarray}  
The coefficients $A$, $B$, $B'$, $C$, $D$, and $E$ can be determined in a fit 
of the experimental excitation energies.

\begin{figure}
\includegraphics[width=8cm]{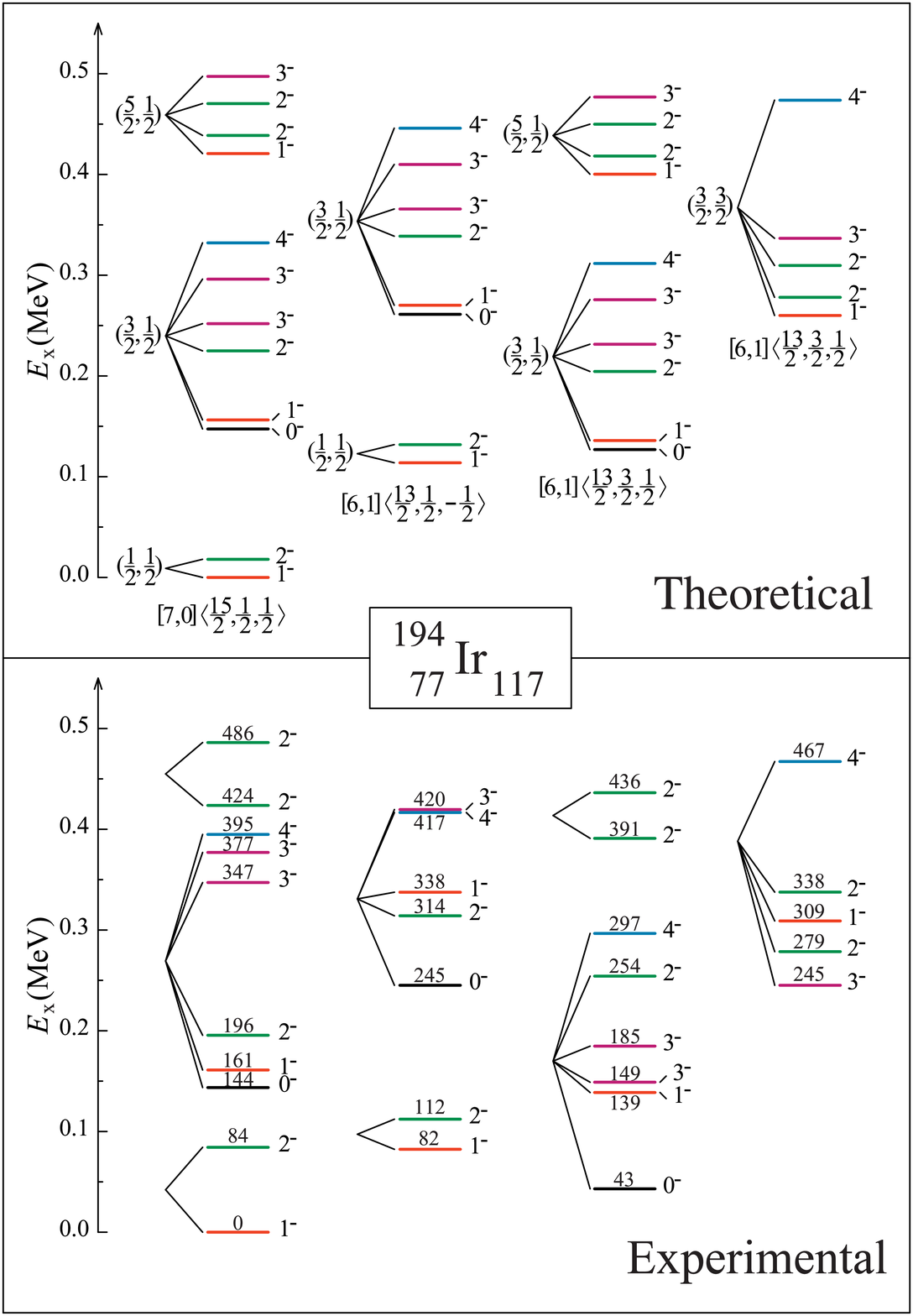} 
\caption{Comparison between the theoretical and experimental spectrum of $^{194}$Ir.}
\label{ir194} 
\end{figure}

The new data from the polarized $(\vec{d},\alpha)$ transfer reaction 
has provided crucial new information about and insight into the structure of the spectrum 
of $^{194}$Ir which led to significant changes in the assignment of levels as compared 
to previous work \cite{joliegarrett}. The main change is that the ground state of 
$^{194}$Ir is now assigned to the band $[N+1,0]$, $(N+\frac{3}{2},\frac{1}{2},\frac{1}{2})$ 
instead of to $[N,1]$, $(N+\frac{1}{2},\frac{1}{2},-\frac{1}{2})$ as in \cite{joliegarrett}. 
In this notation, $N$ is the number of bosons in the odd-odd nucleus ($N=6$ for $^{194}$Ir). 
The new assignment agrees with that of the neighboring odd-odd nucleus $^{196}$Au 
\cite{metz,groeger,wirth}. Fig.~\ref{ir194} shows the negative parity levels of $^{194}$Ir 
in comparison with the theoretical spectrum in which it is assumed that these levels 
originate from the $\nu 3p_{1/2}$, $\nu 3p_{3/2}$, $\nu 2f_{5/2} \otimes \pi 2d_{3/2}$ 
configuration. The theoretical energy spectrum is calculated using the energy formula of 
Eq.~(\ref{npsusy}) with $A+B = 35.0$, $B' = -33.6$, $C = 35.1$, $D = 6.3$, and $E = 4.5$ 
(all in keV). This parameter set is a lot closer to the parameter values used for $^{196}$Au 
\cite{groeger} than the ones in \cite{joliegarrett}, indicating systematics in this zone of 
the nuclear chart. Given the complex nature of the spectrum of heavy odd-odd nuclei, the 
agreement is remarkable. There is an almost one-to-one correlation between the experimental 
and theoretical level schemes \cite{balodis}. 

\begin{figure}
\includegraphics[width=8cm]{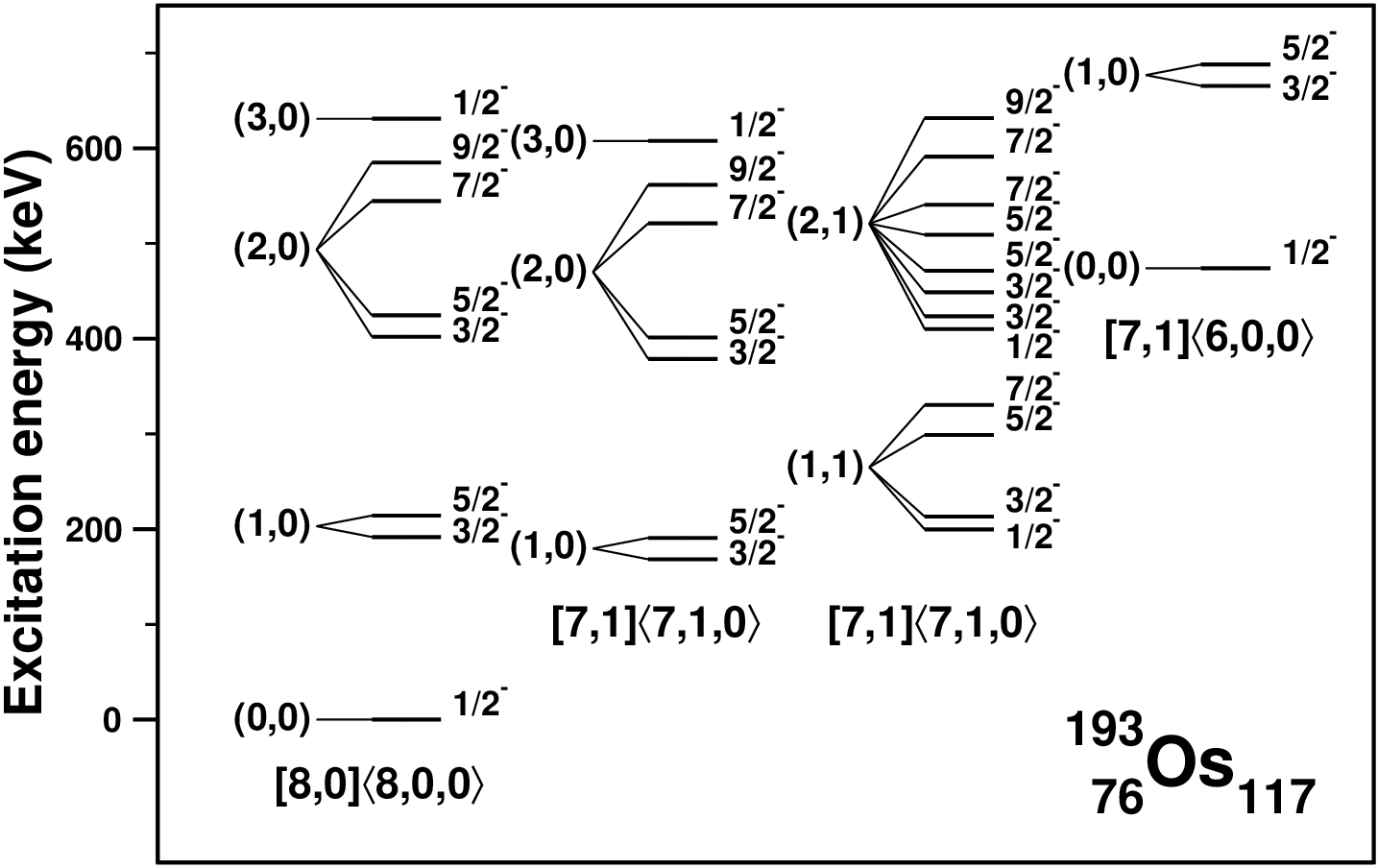} 
\caption{Prediction of the spectrum of $^{193}$Os for the 
$U_{\nu}(6/12) \otimes U_{\pi}(6/4)$ supersymmetry.}
\label{os193} 
\end{figure}

The successful description of the odd-odd nucleus $^{194}$Ir opens the possibility 
of identifying a second quartet of nuclei in the $A \sim 190$ mass region with 
$U(6/12)_{\nu} \otimes U(6/4)_{\pi}$ supersymmetry. The new quartet consists of the 
nuclei $^{192,193}$Os and $^{193,194}$Ir and is characterized by 
${\cal N}_{\pi}=3$ and ${\cal N}_{\nu}=5$. Whereas the $^{192}$Os and $^{193,194}$Ir 
nuclei are well-known experimentally, the available data for $^{193}$Os is rather scarce. 
In Fig.~\ref{os193} we show the predicted spectrum for $^{193}$Os obtained from 
Eq.~(\ref{npsusy}) using the same parameter set as for $^{194}$Ir \cite{balodis}. 
The ground state of $^{193}$Os has spin and parity $J^P=\frac{3}{2}^{-}$, which seems to 
imply that the second band with labels $[7,1]$, $(7,1,0)$ is the ground state band, rather 
than $[8,0]$, $(8,0,0)$. This ordering of bands is supported by preliminary results 
from the one-neutron transfer reaction $^{192}$Os$(\vec{d},p){}^{193}$Os \cite{eisermann}. 

An analysis of the energy spectra of the four nuclei that make up the quartet 
shows that the parameter set obtained in 1981 for the pair $^{192}$Os-$^{193}$Ir
\cite{susy} is very close to that of $^{194}$Ir \cite{balodis}, which indicates  
that the nuclei $^{192,193}$Os and $^{193,194}$Ir may be interpreted in terms of  
a quartet of nuclei with $U(6/12)_{\nu} \otimes U(6/4)_{\pi}$ supersymmetry. 
A more detailed study of the energy spectra of these nuclei is in progress \cite{osir}. 

\subsection{Two-nucleon transfer reactions}

Two-nucleon transfer reactions probe the structure of the final nucleus 
through the exploration of two-nucleon correlations that may be present. 
The spectroscopic strengths not only depend on the similarity between 
the states in the initial and final nucleus, but also on the correlation 
of the transferred pair of nucleons. 

In this section, the recent data on the $^{196}$Pt$(\vec{d},\alpha){}^{194}$Ir  
reaction \cite{balodis} are compared with the predictions from the 
$U_{\nu}(6/12)\otimes U_{\pi}(6/4)$ supersymmetry. This reaction involves the 
transfer of a proton-neutron pair, and hence measures the neutron-proton 
correlation in the odd-odd nucleus. 
The spectroscopic strengths $G_{LJ}$ 
\begin{equation}
G_{LJ} = | \sum_{j_{\nu} j_{\pi}} g_{j_{\nu}j_{\pi}}^{LJ} 
\left< ^{194}\mbox{Ir} \right\| 
( a_{j_{\nu}}^{\dagger} a_{j_{\pi}}^{\dagger} )^{(\lambda)}
\left\| ^{196}\mbox{Pt} \right> |^2 ~, 
\label{glj}
\end{equation}
depend on the reaction mechanism via the coefficients $g_{j_{\nu}j_{\pi}}^{LJ}$ 
and on the nuclear structure part via the reduced matrix elements. 

In order to compare with experimental data we calculate the relative strengths 
$R_{LJ}=G_{LJ}/G_{LJ}^{\textrm{ref}}$, where $G_{LJ}^{\textrm{ref}}$ is the 
spectroscopic strength of the reference state. The ratios of spectroscopic 
strengths to final states with $(\tau_1,\tau_2)=(\frac{3}{2},\frac{1}{2})$ 
provide a direct test of the nuclear wave functions, since they can only be 
excited by a single tensor operator \cite{barea2}. In Table~\ref{tab:ratios} we 
show the ratios for different final states with 
$(\tau_1,\tau_2)=(\frac{3}{2},\frac{1}{2})$. 

\begin{table}
\centering
\caption{\label{tab:ratios} Ratios of spectroscopic strengths for $(\vec{d},\alpha)$ 
reactions $R_{LJ}(\rm ee \rightarrow oo)$ to final states with 
$(\tau_1,\tau_2)=(\frac{3}{2},\frac{1}{2})$. $N$ is the number of bosons in the 
odd-odd nucleus ($N=6$ for $^{196}$Pt $\rightarrow$ $^{194}$Ir).}
\begin{tabular*}{86mm}{c@{\extracolsep\fill}ccc} 
\hline\\[-2mm]
$[N_1,N_2]$ & $(\Sigma_1,\Sigma_2,\Sigma_3)$ & $(\sigma_1,\sigma_2,\sigma_3)$ & $R_{LJ}$ \\[1mm] 
\hline\\[-2mm]
$[N,1]$ & $(N,1,0)$ & $(N+\frac{1}{2},\frac{3}{2},\frac{1}{2})$ & 1 \\[1mm] 
$[N,1]$ & $(N,1,0)$ & $(N+\frac{1}{2},\frac{1}{2},-\frac{1}{2})$  &  
$\frac{N+4}{15N}$  \\[1mm] 
$[N,1]$ & $(N,1,0)$ & $(N-\frac{1}{2},\frac{3}{2},-\frac{1}{2})$  &  
$\frac{(N+4)(N+1)(N-1)}{N(N+3)(N+5)}$ \\[1mm] 
$[N,1]$ & $(N,1,0)$ & $(N-\frac{1}{2},\frac{1}{2},\frac{1}{2})$  & 
$\frac{(N+1)(N-1)}{15(N+3)(N+5)}$ \\[1mm] 
$[N+1,0]$ & $(N+1,0,0)$ & $(N+\frac{3}{2},\frac{1}{2},\frac{1}{2})$  &
$\frac{2(N+4)(N+6)}{15N(N+3)}$  \\[1mm] 
$[N+1,0]$ & $(N+1,0,0)$ & $(N+\frac{1}{2},\frac{1}{2},-\frac{1}{2})$  &  
$\frac{2(N+4)}{15(N+3)}$  \\[1mm] 
\hline
\end{tabular*}
\end{table}

\begin{figure}
\begin{center}
\includegraphics[width=8cm]{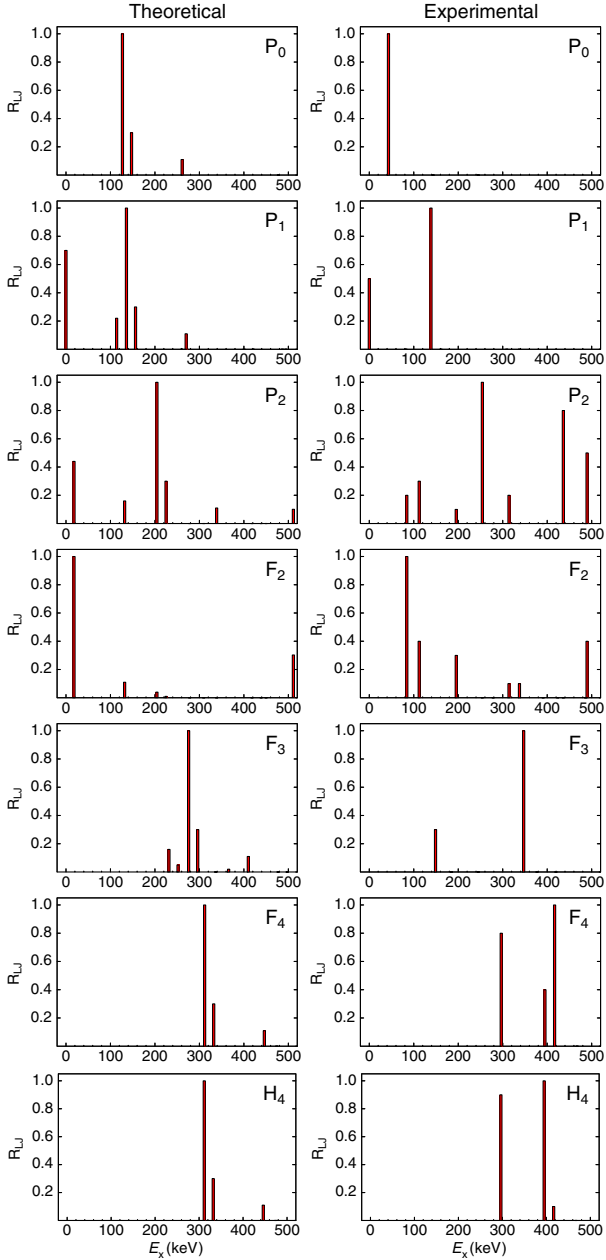} 
\caption{\label{ratios} Comparison of theoretical (left panels) and experimental 
values (right panels) of ratios $R_{LJ}$ of spectroscopic strengths.}
\end{center}
\end{figure}

Fig.~\ref{ratios} shows the ratios of spectroscopic strengths of two-nucleon transfer 
reactions $R_{LJ}$ compared with the theoretical predictions from nuclear supersymmetry.  
The reference states are easily identified, since they are normalized to one. 
The calculations were carried out without the introduction of any new parameter since  
the coefficients $g_{j_{\nu}j_{\pi}}^{LJ}$ appearing in the transfer operator of Eq.~(\ref{glj}) 
were taken from the study of the $^{198}$Hg$(\vec d,\alpha)^{196}$Au reaction \cite{barea2}. 
In general, there is good overall agreement between the experimental and 
theoretical values, especially if we take into account the simple form of the 
operator in the calculation of the two-nucleon transfer reaction intensities. 
The deviations observed in the $P_2$ and $F_2$ transfers are most likely due to single-particle 
configurations outside the model space. For the $P_0$, $P_1$, and $F_3$ distributions
the experimental ($\vec d,\alpha$) detection limits for weakly populated 0$^-$, 1$^-$, 
and 3$^-$ states prevent a better agreement. 

\section{Correlations}

The nuclei belonging to a supersymmetric quartet are described by a single 
Hamiltonian, and hence the wave functions, transition and transfer rates are 
strongly correlated. As an example of these correlations, we consider here the 
transfer reactions between the $^{194,195}$Pt and $^{192,193}$Os nuclei. 
The Pt and Os nuclei are connected by one-neutron transfer reactions within 
the same supersymmetric quartet $^{194}$Pt $\leftrightarrow$ $^{195}$Pt and 
$^{192}$Os $\leftrightarrow$ $^{193}$Os, whereas the transitions between the Pt 
and Os nuclei involve the transfer of a proton pair between different quartets 
$^{194}$Pt $\leftrightarrow$ $^{192}$Os and $^{195}$Pt $\leftrightarrow$ $^{193}$Os. 

\subsection{Generalized F-spin}

The correlations between different transfer reactions can be derived in an 
elegant and explicit way by a generalization of the concept of $F$-spin 
which was introduced in the neutron-proton IBM \cite{otsuka} in order to 
distinguish between proton and neutron bosons. 
 
The eigenstates of the $U(6/12)_{\nu} \otimes U(6/4)_{\pi}$ supersymmetry are 
characterized by the irreducible representations $[N_1,N_2,N_3]$ of $U^{BF_{\nu}}(6)$ 
which arise from the coupling of three different $U(6)$ representations, 
$[N_{\nu}]$ for the neutron bosons, $[N_{\pi}]$ for the proton bosons 
and $[N_{\rho}]$ for the pseudo-orbital angular momentum of the odd neutron 
($N_{\rho}=0$ for the even-even and odd-proton nucleus of the quartet, and $N_{\rho}=1$ 
for the odd-neutron and the odd-odd nucleus). In analogy with the three quark flavors 
in the quark model ($u$, $d$ and $s$), also here we have three different types 
of identical objects ($\pi$, $\nu$ and $\rho$), which can be distinguished by $F$-spin 
and hypercharge $Y$. The two kinds of bosons form an $F$-spin doublet, $F=\frac{1}{2}$,  
with charge states $F_z=\frac{1}{2}$ for protons ($\pi$) and $F_z=-\frac{1}{2}$ for 
neutrons ($\nu$) \cite{otsuka}. In the framework of the generalized $F$-spin, we 
assign in addition a hypercharge quantum number to the bosons $Y=\frac{1}{3}$. 
The pseudo-orbital part ($\rho$) has $F=F_z=0$ and $Y=-\frac{2}{3}$. 

Group theoretically, the generalized $F$-spin is defined by the reduction 
\begin{eqnarray}
\begin{array}{ccccc} U(18) &\supset& U(6) &\otimes& U(3) \\
\downarrow && \downarrow && \downarrow \\
\, [N] && [N_1,N_2,N_3] && [N_1,N_2,N_3] 
\label{fspin}
\end{array}
\end{eqnarray}
Here $U(6)$ is to be identified with the $U^{BF_{\nu}}(6)$ of the group reduction 
of Eq.~(\ref{chain}), which is the result of first coupling the bosons at the level 
of $U(6)$ followed by coupling the orbital part  
\begin{eqnarray}
\left| [N_{\nu}],[N_{\pi}];[N_{\nu}+N_{\pi}-i,i],[N_{\rho}];[N_1,N_2,N_3] \right> ~.
\end{eqnarray}
This sequence of $U(6)$ couplings can be described in a completely equivalent 
way by the three-dimensional index group $U(3)$ of Eq.~(\ref{fspin}) which can 
be reduced to
\begin{eqnarray}
\begin{array}{ccccccccc} U(3) &\supset& SU(3) &\supset& [SU(2) &\supset& SO(2)] &\otimes& U(1) \\
&& \downarrow && \downarrow && \downarrow && \downarrow \\
&& (\lambda,\mu) && F && F_z && Y 
\end{array}
\end{eqnarray}
The relation between the two sets of quantum numbers is given by
\begin{eqnarray}
(\lambda,\mu) &=& (N_1-N_2,N_2-N_3) ~,
\nonumber\\
F &=& \frac{1}{2} \left( N_{\pi}+N_{\nu}-2i \right) ~,
\nonumber\\
F_z &=& \frac{1}{2} \left( N_{\pi}-N_{\nu} \right) ~,
\nonumber\\
Y &=& \frac{1}{3} \left( N_{\pi}+N_{\nu}-2N_{\rho} \right) ~.
\end{eqnarray}
As a result, matrix elements between states with the same quantum numbers 
but different $U(6)$ couplings are then related by $SU(3)$ isoscalar factors 
(or Clebsch-Gordan coefficients for $SU(3)$), and hence correlations between 
different transfer reactions can be derived in terms of these isoscalar 
factors by means of the concept of generalized $F$-spin. 

\subsection{One-neutron transfer} 

In a study of the $^{194}$Pt $\rightarrow$ $^{195}$Pt stripping reaction 
it was found \cite{bi} that one-neutron $j=3/2$, $5/2$ transfer 
reactions can be described by the operator 
\begin{eqnarray}
P_{\nu}^{(j) \, \dagger} &=& \frac{\alpha_j}{\sqrt{2}} \left[ 
\left( \tilde{s}_{\nu} \times a^{\dagger}_{\nu,j} \right)^{(j)} - 
\left( \tilde{d}_{\nu} \times a^{\dagger}_{\nu,\frac{1}{2}} \right)^{(j)} \right] ~. 
\end{eqnarray}
It is convenient to take ratios of intensities, since they do not depend on the 
value of the coefficient $\alpha_j$ and hence provide a direct test of the wave 
functions. For the stripping reaction $^{194}$Pt $\rightarrow$ $^{195}$Pt 
(ee $\rightarrow$ on) the ratio of intensities for the excitation of the 
$(\tau_1,\tau_2)=(1,0)$, $L=2$ doublet with $J=3/2$, $5/2$ belonging to the first 
excited band with $[N+1,1]$, $(N+1,1,0)$ relative to that of the ground state band 
$[N+2]$, $(N+2,0,0)$ is given by \cite{bi} 
\begin{eqnarray}
R({\rm ee \rightarrow on}) = \frac{(N+1)(N+3)(N+6)}{2(N+4)} ~,
\label{ratio}
\end{eqnarray}
which gives $R=29.3$ for $^{194}$Pt $\rightarrow$ $^{195}$Pt ($N=5$), 
to be compared to the experimental value of 19.0 for $j=5/2$, and 
$R=37.8$ for $^{192}$Os $\rightarrow$ $^{193}$Os ($N=6$).   
The equivalent ratio for the inverse pick-up reaction is given by
\begin{eqnarray}
R({\rm on \rightarrow ee}) = R({\rm ee \rightarrow on}) 
\frac{N_{\pi}+1}{(N+1)(N_{\nu}+1)} ~. 
\label{corr}
\end{eqnarray}
which gives $R=1.96$ for $^{195}$Pt $\rightarrow$ $^{194}$Pt ($N_{\pi}=1$ and 
$N_{\nu}=4$) and $R=3.24$ for $^{193}$Os $\rightarrow$ $^{192}$Os ($N_{\pi}=2$ 
and $N_{\nu}=4$). This means that the mixed symmetry $L=2$ state is predicted 
to be excited more strongly than the first excited $L=2$ state.  

This correlation between pick-up and stripping reactions has been derived in a general 
way only using the symmetry relations that exist between the wave functions of the 
even-even and odd-neutron nuclei of the supersymmetric quartet. The factor 
in the right-hand side of Eq.~(\ref{corr}) is the result of a ratio of two $SU(3)$  
isoscalar factors. It is important to emphasize, that Eqs.~(\ref{ratio}) and 
(\ref{corr}) are parameter-independent predictions which are a direct consequence 
of nuclear SUSY and which can be tested experimentally.  

\subsection{Two-proton transfer} 

The two supersymmetric quartets in the mass $A \sim 190$ region differ by two protons. 
In principle, the connection between the two quartets can be studied by two-proton 
transfer reactions. In the IBM, two-proton transfer operator is, in first order, given by 
\ba
P_{\pi}^{\dagger} = \alpha \, s_{\pi}^{\dagger} ~, \hspace{1cm}
P_{\pi} = \alpha \, s_{\pi} ~.
\ea
Whereas the operator $s_{\pi}$ only excites the ground state of the final nucleus, 
$s_{\pi}^{\dagger}$ can also populate excited states. 

\begin{table}
\centering
\caption{\label{ree} Ratios of spectroscopic strengths for two-proton 
transfer reactions between even-even nuclei $R(\rm ee \rightarrow ee)$ 
to final states with $(\tau_1,\tau_2)=(0,0)$. 
$N$ is the number of bosons in the odd-odd nucleus of the same quartet as the 
initial even-even nucleus ($N=5$ for $^{194}$Pt $\rightarrow$ $^{192}$Os).}
\begin{tabular*}{86mm}{c@{\extracolsep\fill}ccc} 
\hline\\[-2mm]
$n$ & $[N_1,N_2]$ & $(\Sigma_1,\Sigma_2,\Sigma_3)$ & $R_n$ \\[1mm] 
\hline\\[-2mm]
1 & $[N+3,0]$ & $(N+3,0,0)$ & $1$ \\[1mm]
2 & $[N+3,0]$ & $(N+1,0,0)$ & $\frac{(N+2)(N+5)}{(N+3)^2(N+6)}$ \\[1mm]
3 & $[N+2,1]$ & $(N+1,0,0)$ & $\frac{(N_{\nu}+1)(N+1)(N+4)}{(N_{\pi}+2)(N+3)^2(N+6)}$ \\[1mm]
\hline
\end{tabular*}
\end{table}

In Table~\ref{ree}, we show 
the results for ratios of spectroscopic strengths between even-even nuclei. 
The selection rules of the operator $s_{\pi}^{\dagger}$ allow the excitation of 
states with with $(\tau_1,\tau_2)=(0,0)$ and $L=0$ belonging to the ground band 
$(\Sigma_1,\Sigma_2,\Sigma_3)=(N+3,0,0)$ and excited bands with $(N+1,0,0)$.  
The corresponding ratios for the odd-neutron nuclei are strongly correlated to those 
of the even-even nuclei (see Tables~\ref{ree} and \ref{ron}) 
\ba
S_{2}({\rm on \rightarrow on}) &=& R_{2}({\rm ee \rightarrow ee}) ~,
\nonumber\\
S_{3a}({\rm on \rightarrow on}) &=& R_{3}({\rm ee \rightarrow ee}) 
\frac{N_{\pi}+2}{(N_{\nu}+1)(N+2)} ~,
\nonumber\\
S_{3b}({\rm on \rightarrow on}) &=& R_{3}({\rm ee \rightarrow ee}) 
\frac{N_{\nu}(N+3)}{(N_{\nu}+1)(N+2)} ~.
\ea
As before, the coefficients in the right-hand side correspond to the ratio of two 
$SU(3)$ Clebsch-Gordan coefficients. 

\begin{table}
\centering
\caption{\label{ron} Ratios of spectroscopic strengths for two-proton 
transfer reactions between odd-neutron nuclei $S(\rm on \rightarrow on)$ 
to final states with $(\tau_1,\tau_2)=(0,0)$.  
$N$ is the number of bosons in the odd-odd nucleus of the same quartet as the 
initial odd-neutron nucleus ($N=5$ for $^{195}$Pt $\rightarrow$ $^{193}$Os).}
\begin{tabular*}{86mm}{c@{\extracolsep\fill}cccc} 
\hline\\[-2mm]
$n$ & $[N+2-i,i]$ & $[N_1,N_2]$ & $(\Sigma_1,\Sigma_2,\Sigma_3)$ & $S_n$ \\[1mm] 
\hline\\[-2mm]
1 & $[N+2,0]$ & $[N+3,0]$ & $(N+3,0,0)$ & $1$ \\[1mm]
2 & $[N+2,0]$ & $[N+3,0]$ & $(N+1,0,0)$ & $R_2$ \\[1mm]
3a & $[N+2,0]$ & $[N+2,1]$ & $(N+1,0,0)$ & $R_3 \frac{N_{\pi}+2}{(N_{\nu}+1)(N+2)}$ \\[1mm]
3b & $[N+1,1]$ & $[N+2,1]$ & $(N+1,0,0)$ & $R_3 \frac{N_{\nu}(N+3)}{(N_{\nu}+1)(N+2)}$ \\[1mm]
\hline
\end{tabular*}
\end{table}

\section{Summary and conclusions}

In conclusion, in this contribution we have presented evidence for the existence of a second 
quartet of nuclei in the $A\sim 190$ region with $U(6/12)_{\nu} \otimes U(6/4)_{\pi}$ 
supersymmetry, consisting of the $^{192,193}$Os and $^{193,194}$Ir nuclei. The 
analysis is based on new experimental information on $^{194}$Ir. In particular,  
the $(\vec{d},\alpha)$ reaction is important to establish the spin and parity 
assignments of the energy levels, and to provide insight into the structure of 
the spectrum of $^{194}$Ir. Given the complexity of the $A \sim 190$ mass region, 
the simple yet detailed description of $^{194}$Ir in a supersymmetry scheme is 
remarkable.  

Nuclear supersymmetry establishes precise links among the spectroscopic properties 
of different nuclei. This relation has been used to predict the energies of 
$^{193}$Os. Since the wave functions of the members of a supermultiplet are connected 
by symmetry, there exists a high degree of correlation between different one- and 
two-nucleon transfer reactions not only between nuclei belonging to the same quartet, 
but also for nuclei from different multiplets. 
As an example, we studied the correlations between one-neutron transfer reactions 
and two-proton transfer reactions. In the former case, nuclear supersymmetry 
predicts that the $L=2$ mixed symmetry states in the even-even nuclei $^{194}$Pt 
and $^{192}$Os are excited much stronger (two to three times as strong) than the 
first excited $L=2$ state.

In order to establish the existence of a second supersymmetric quartet of nuclei 
in the $A \sim 190$ mass region, it is crucial that the nucleus $^{193}$Os be studied 
in more detail experimentally. The predictions for correlations between one-neutron 
transfer reactions in Pt and Os can be tested experimentally by combining for example 
$(\vec{d},p)$ stripping and $(p,d)$ pick-up reactions.  

\section*{Acknowledgments}

This work was supported in part by PAPIIT-UNAM (grant IN113808).

\end{document}